\UseRawInputEncoding
\pdfoutput=1
\documentclass[conference]{IEEEtran}
\IEEEoverridecommandlockouts
\usepackage{cite}
\usepackage{amsmath,amssymb,amsfonts}
\usepackage{url}
\usepackage{graphicx}
\usepackage{textcomp}
\usepackage{xcolor}
\usepackage[utf8]{inputenc}
\usepackage[hidelinks]{hyperref}
\usepackage{orcidlink}
\usepackage{tikz}
\definecolor{medium_blue}{rgb}{0.0, 0.4, 0.8}

\newcommand\copyrighttext{%
  \footnotesize
  \textcopyright\ 2025 IEEE.
  Personal use of this material is permitted.
  Permission from IEEE must be obtained for all other uses, in any current or future media, including reprinting/republishing this material for advertising or promotional purposes, creating new collective works, for resale or redistribution to servers or lists, or reuse of any copyrighted component of this work in other works.\\
}
\newcommand\copyrightnotice{%
  \begin{tikzpicture}[remember picture,overlay]
    \node[anchor=south,yshift=25pt] at (current page.south)
      {\parbox{\dimexpr\textwidth-\fboxsep-\fboxrule\relax}{\copyrighttext}};
  \end{tikzpicture}%
}

\begin{document}

\title{FAIR GraphRAG: A Retrieval-Augmented Generation Approach for Semantic Data Analysis\\
}

\author{\IEEEauthorblockN{Marlena Flüh\,\orcidlink{0009-0008-6379-2988}}
\IEEEauthorblockA{\textit{Data Stream Management and Analysis} \\
\textit{RWTH Aachen University}\\
Aachen, Germany \\
marlena.flueh@rwth-aachen.de}
\and
\IEEEauthorblockN{Soo-Yon Kim\,\orcidlink{0000-0001-5975-0031}}
\IEEEauthorblockA{\textit{Data Stream Management and Analysis} \\
\textit{RWTH Aachen University}\\
Aachen, Germany \\
kim@dbis.rwth-aachen.de}
\and
\IEEEauthorblockN{Carolin Victoria Schneider\,\orcidlink{0000-0002-6728-9246}}
\IEEEauthorblockA{\textit{Gastroenterology, Metabolic Diseases} \\
\textit{and Intensive Care} \\
\textit{University Hospital RWTH Aachen}\\
Aachen, Germany \\
cschneider@ukaachen.de}
\and
\IEEEauthorblockN{Sandra Geisler\,\orcidlink{0000-0002-8970-6282}}
\IEEEauthorblockA{\textit{Data Stream Management and Analysis} \\
\textit{RWTH Aachen University}\\
Aachen, Germany \\
geisler@dbis.rwth-aachen.de}
}

\maketitle
\copyrightnotice
\begin{abstract}
Retrieval-Augmented Generation (RAG) addresses the limitations of Large Language Models (LLMs) when providing responses to domain-specific questions. Graph-based RAG approaches, such as GraphRAG, enhance retrieval by capturing semantic relationships within knowledge graphs (KGs). 
While the FAIR principles (\textit{Findability, Accessibility, Interoperability, and Reusability}) are becoming prevalent for scientific data management, especially in complex domains such as medicine, 
existing RAG approaches lack a structured FAIRification of the underlying knowledge resources. 
This lack limits their potential for FAIR information retrieval in these domains.
To address this gap, we introduce FAIR GraphRAG, a novel framework that integrates FAIR Digital Objects (FDOs) as the fundamental units of a graph-based retrieval system. Each graph node represents an FDO that incorporates core data, metadata, persistent identifiers, and semantic links. We leverage LLMs to support schema construction and automated extraction of content and metadata from data sources. The framework was co-designed by physicians and computer scientists to ensure technical and clinical relevance.
We apply FAIR GraphRAG to a biomedical dataset in gastroenterology, demonstrating its applicability to RNA-sequencing data. 
Beyond ensuring adherence to the FAIR principles, FAIR GraphRAG significantly improves question answering accuracy, coverage, and explainability, particularly for complex queries involving metadata and ontology links.
This work shows the feasibility of combining FAIR data practices with graph-based retrieval techniques. We see potential for applying our approach to other specialized fields such as education and business. 
\end{abstract}


\section{Introduction}

Imagine a clinician asking: “Does the patient's diagnosis require dose adjustment for prescribed medication?” --- a query that requires integrating patient records with standardized drug databases. In practice, patient diagnoses often use custom terms, while drug information relies on standardized vocabularies such as SNOMED CT\footnote{\url{https://www.nlm.nih.gov/healthit/snomedct/index.html}}. Without linking patient data to standard ontologies, even advanced AI systems fail to match diagnoses to correct dosing recommendations. This mismatch leads to incomplete or incorrect results and can compromise patient care.

Addressing such challenges requires sophisticated methods for extracting and connecting knowledge from different data sources. Recent advances in Large Language Models (LLMs) within natural language processing have revolutionized the extraction of information from diverse sources such as text and tabular data~\cite{pan_large_2023}. 
These advancements impact various fields including healthcare, finance and education~\cite{peng_graph_2024, collarana_graphrag_2025}.
However, despite their capabilities, LLMs lack domain-specific knowledge, when it is not part of the LLMs' training corpus. This knowledge gap can lead to hallucination or factually incorrect outputs~\cite{peng_graph_2024, yu_evaluation_2025}.

This is why recent studies have focused on Retrieval-Augmented Generation (RAG)~\cite{zhao_retrieval-augmented_2024, peng_graph_2024, collarana_graphrag_2025}, which supplements LLMs with external knowledge to improve response accuracy. RAG integrates domain-specific sources, such as datasets, enhancing the retrieval of factual information and reducing hallucination. In medicine, RAG systems can answer questions by retrieving linked records such as patient diagnoses, disease terms, and their relationships to medication guidelines. The performance of RAG systems is highly dependent on the quality and relevance of the retrieved information, underscoring the importance of the underlying knowledge source.

Knowledge Graphs (KGs) are semantically rich, interconnected structures designed to represent relationships between entities in a machine-readable format~\cite{ji_survey_2022}. 
This data structure facilitates understanding and discovering underlying patterns, making KGs particularly useful in domains requiring an organized, comprehensive view of complex information. 
Moreover, KGs enable the integration of diverse data sources, making them well-suited for complex analytical tasks~\cite{collarana_graphrag_2025}. A RAG system that retrieves relevant information from graph databases can be seen as 
a subcategory of RAG called GraphRAG~\cite{peng_graph_2024}. While KGs provide factual knowledge, LLMs remain essential for interpreting complex queries and generating context-aware answers~\cite{collarana_graphrag_2025}.

Returning to the opening example, fully answering clinical queries requires a standardized way to incorporate all relevant information---such as ontology links---into the underlying knowledge resources, ensuring accurate and interoperable retrieval. The FAIR principles offer widely recognized guidelines for effective data management~\cite{de_smedt_fair_2020}. While initial efforts have produced FAIR KGs that include rich metadata, standard vocabularies, persistent identifiers (PIDs), and provide  human- and machine-accessible interfaces~\cite{brandizi_towards_2018, pestryakova_covidpubgraph_2022, sherif_ingridkg_2023, vogt_fair_2023}, to our knowledge, there is no comprehensive solution for the structured FAIRification of the knowledge underlying RAG systems. As the opening example illustrates, addressing this gap is critical to ensure that information from heterogeneous sources can be efficiently discovered, linked, accessed, and reused in AI-driven systems.

\begin{figure}[t]
\centerline{\includegraphics[width=0.2\textwidth]{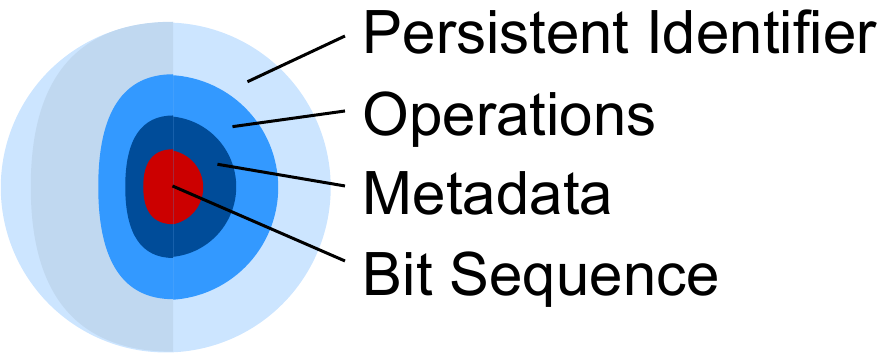}}
\caption{A High-level FAIR Digital Object model. Adopted from Wittenburg et al.~\cite{wittenburg_digital_2019}}
\label{fig:fdo_blue}
\end{figure}

Building on prior work in GraphRAG and FAIR KGs, 
we introduce a novel FAIR GraphRAG framework that uses FAIR Digital Objects (FDOs), visualized in 
Fig.~\ref{fig:fdo_blue}, as the core knowledge units within the retrieval system. 
Each node in our graph represents an FDO that fully aligns with the FAIR principles by incorporating 
in its "shells" a data artifact at the core, metadata describing the artifact, and a PID identifying it uniquely.
While implementing GraphRAG-based data exploration, the framework itself is also LLM-based --- we evaluate the usage of LLMs for the creation of FAIR KG construction, including tasks such as schema design and automated extraction of entities and metadata from datasets. We use the term \textit{dataset} for structured and unstructured data collections.
Our main contributions are as follows: (1) we introduce the FAIR GraphRAG framework and formalize a FAIR KG model based on FDOs; (2) we develop an end-to-end pipeline for constructing FAIR KGs; and (3) we demonstrate the practical value of our approach through implementation and evaluation on a biomedical dataset. The framework was developed in collaboration by physicians and computer scientists.

The paper is structured as follows. In Section~\ref{sec:relatedwork}, we introduce the FAIR principles and the concept of FDOs, and discuss state-of-the-art GraphRAG approaches, highlighting the lack of structured FAIRification in existing methods. In Section~\ref{sec:method}, we formalize, construct, and implement a FAIR KG based on FDOs within a FAIR GraphRAG framework. Section~\ref{sec:experiment} presents an evaluation of our approach on a biomedical dataset, demonstrating its applicability. Finally, Section~\ref{sec:conclusion} summarizes our contributions and outlines directions for future work.

\section{Related Work}
\label{sec:relatedwork}
This section gives an overview of concepts and recent developments relevant to our work. 
We begin by discussing the FAIR principles and examine their application to digital objects in the form of FDOs. 
Next, we review approaches to construct KGs adhering to the FAIR principles, including efforts to integrate FDOs as semantically meaningful units within graph structures. 
We highlight recent progress in using LLMs for KG construction and note that existing methods do not yet incorporate the FAIR principles or FDOs. 
Finally, we discuss the emergence of RAG and its extension to graph-based retrieval, with a focus on domain-specific applications.

\subsection{FAIR Digital Objects}
The FAIR principles were introduced to improve the management, sharing, and reuse of scientific data, making the data \textit{Findable, Accessible, Interoperable}, and \textit{Reusable}~\cite{wilkinson_fair_2016}. 
To implement these principles, the concept of FDOs was developed~\cite{schwardmann_digital_2020}.
An FDO is a standardized digital entity designed to be understandable and actionable by humans and machines~\cite{de_smedt_fair_2020}. It brings together all the information and functionalities needed to work with digital data objects in a consistent way. As illustrated in Fig.~\ref{fig:fdo_blue}, an FDO consists of several core elements, shown here on the example of a published research article:
\begin{itemize}
    \item \textbf{Bit sequence:} The core data in digital format (e.g., the PDF file of the article).
    \item \textbf{Metadata:} Information describing the data, such as title, authors, publication date, and journal name.
    \item \textbf{Operations:} Standardized, machine-readable actions inspired by object-oriented programming (e.g., \texttt{get\_metadata}, \texttt{download}).
    \item \textbf{Persistent Identifier (PID):} Permanent unique identifier for reliable access (e.g., \texttt{doi:10.2345/\allowbreak article123}).
\end{itemize}
By packaging data in this way, FDOs support automation, reproducibility, and interoperability, making scientific data more valuable and reusable. Their practical relevance is demonstrated by applications in organizational data governance~\cite{queralt-rosinach_applying_2022, vogt_towards_2024} and biomedical domains~\cite{flynn_linked_2022}. 

\subsection{FAIR Principles and Knowledge Graphs}
Constructing high-quality KGs from heterogeneous data remains a foundational challenge~\cite{cappiello_52_2024}.
Several publications have addressed the creation of graph structures that adhere to the FAIR principles, ranging from FAIR biological knowledge networks, over a FAIR COVID-19 KG, to a FAIR graffiti KG~\cite{brandizi_towards_2018, pestryakova_covidpubgraph_2022, sherif_ingridkg_2023}. While these works focus on making the entire KG FAIR by assigning a PID, metadata, and standardized access, they leave open the challenge of identifying and FAIRifying semantically meaningful subunits within the graph. 
Such subunits enable fine-grained access control, efficient data integration, automated discovery and reuse by machines, increasing the utility and interoperability of KGs.
Some approaches further (1) structure KGs into semantic subgraphs~\cite{vogt_semantic_2024} or (2) link metadata for FDOs~\cite{flynn_linked_2022}.
While creating KGs based on FDOs has been explored~\cite{vogt_semantic_2024}, existing work focuses mainly on linked metadata and semantic units, without addressing aspects such as linking core content, ontology term mapping, or the use of LLMs for graph construction.

\subsection{LLMs for Knowledge Graph Construction} Recent advances demonstrate the use of LLMs for constructing knowledge graphs from heterogeneous and unstructured data sources~\cite{sun_docs2kg_2024, lairgi_itext2kg_2024, zhang_extract_2024}. Approaches range from building unified KGs across dataset types to incremental and schema-flexible construction directly from text. LLMs are leveraged for key tasks such as schema design, entity and relation extraction, and automated enrichment of graph content. However, current LLM-driven KG construction approaches do not address the FAIR principles or the use of FDOs. Closing this gap is crucial, as ensuring KGs are FAIR enhances data findability, accessibility, interoperability, and reusability, making KGs more useful for both humans and machines.

\subsection{Graph-Retrieval Augmented Generation} RAG improves response generation by retrieving relevant information from available data sources~\cite{zhao_retrieval-augmented_2024}. When using LLMs, the relevant information is added to the LLM context window, which is the span of text the LLM can process at once, along with the original user question, enabling the LLM to generate a more informed answer.
Hence, RAG allows the retrieval of domain-specific information, minimizing issues such as hallucination or factually incorrect outputs~\cite{peng_graph_2024, yu_evaluation_2025}. However, traditional RAG often overlooks the inherent relationships within the underlying retrieved text, failing to represent its interconnected structure~\cite{peng_graph_2024}. Retrieving knowledge from graph structures (e.g., nodes and their interconnections) allows us to consider the semantic relationships between concepts or entities mentioned in the text, enhancing their ability to produce context-aware and factually accurate responses~\cite{peng_graph_2024}. 
KG-based LLM summarization techniques can be used to answer global questions over an entire text corpus, such as “What are the main themes in the dataset?”. This approach consistently outperforms traditional RAG in comprehensiveness and diversity, as shown by evaluation in two datasets~\cite{edge_local_2024}.
A core innovation is the construction of multiple subgraphs by grouping closely-related entities into communities and introducing multiple levels~\cite{edge_local_2024}. In domain-specific settings such as medicine, GraphRAG approaches address the unique challenges posed by the volume, complexity, and sensitivity of heterogeneous medical data, including structured formats (such as lab values), unstructured text (such as free-text clinical notes), and imaging or genomic data. This diversity complicates data integration and semantic mapping.
Recent frameworks link user-provided data to controlled vocabularies, external biomedical knowledge sources, as well as to ontologies such as the Unified Medical Language System (UMLS)~\footnote{\url{https://www.nlm.nih.gov/research/umls/index.html}}~\cite{queralt-rosinach_applying_2022}, thereby also supporting the FAIR principles.  
\newline

Existing RAG approaches incorporate elements that enhance FAIRness, such as multi-level graph structures~\cite{edge_local_2024} to improve findability and accessibility, and the integration of ontologies or controlled vocabularies~\cite{brandizi_towards_2018} to support interoperability and reusability. Additionally, some efforts in the literature have focused on constructing FAIR KGs, with some of them leveraging linked FDOs as basic units within graph structures~\cite{vogt_semantic_2024, flynn_linked_2022}.
However, these approaches have limitations: none have investigated the automated construction of FAIR KGs, nor have they proposed a systematic FAIRification of the underlying data resources of RAG frameworks. This leaves a gap in FAIR KG construction and the potential for enhanced retrieval quality within RAG pipelines.

As demonstrated with the opening example, linking data to standard ontologies as part of FAIRification leads to measurable improvements in information retrieval performance. Systematically integrating FAIR concepts holds significant potential for the development of more transparent, reusable, and reliable RAG systems. Building on these observations, we introduce a framework that systematically applies the FAIR principles and leverages LLMs for KG construction and user interaction via natural language. The following section details the methodology of our FAIR GraphRAG framework.

\section{Methodology}
\label{sec:method}
\begin{figure}[b]
\centerline{\includegraphics[width=0.49\textwidth]{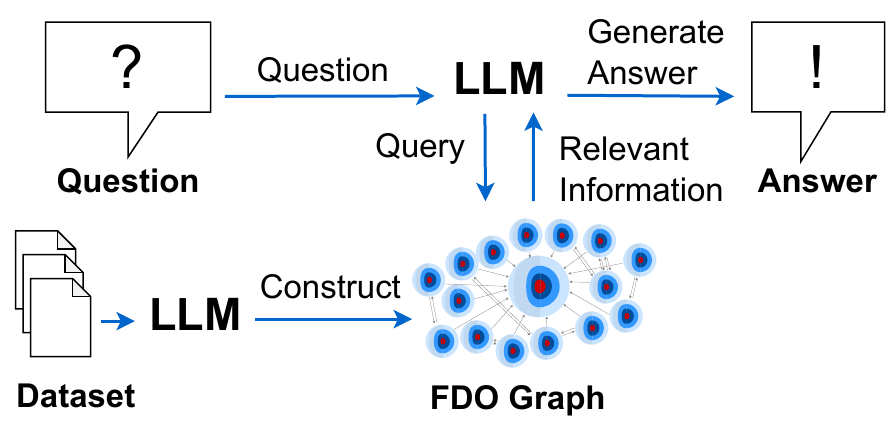}}
\caption{Overview of the FAIR GraphRAG framework. The underlying knowledge graph is composed of nodes, each representing a FAIR Digital Object.}
\label{fig:rag_overview}
\end{figure}
We propose a novel FAIR GraphRAG framework based on a KG of FDOs for analyzing domain-specific datasets, as visualized in Figure~\ref{fig:rag_overview}. 
Most of the existing FAIR KGs~\cite{sherif_ingridkg_2023, pestryakova_covidpubgraph_2022, brandizi_towards_2018} are only FAIR if they are considered as a whole. 
In our approach, each node itself is FAIR as it is modeled as an FDO, i.e., the KG consists of interconnected FDOs. 
We go beyond existing approaches that explore the idea of an FDO network~\cite{flynn_linked_2022, vogt_fair_2023} by (1) linking metadata or core data of FDOs, (2) connecting FDOs to existing ontologies, (3) including multiple node types, (4) supporting the construction of FAIR KGs based on FDOs with LLMs, and (5) enhancing accessibility by using an LLM to answer questions in natural language.
In the following, we derive a formal model of a FAIR FDO KG, describe its construction process, and discuss its integration into a GraphRAG framework.
\begin{figure}[t]
\centerline{\includegraphics[width=0.5\textwidth]{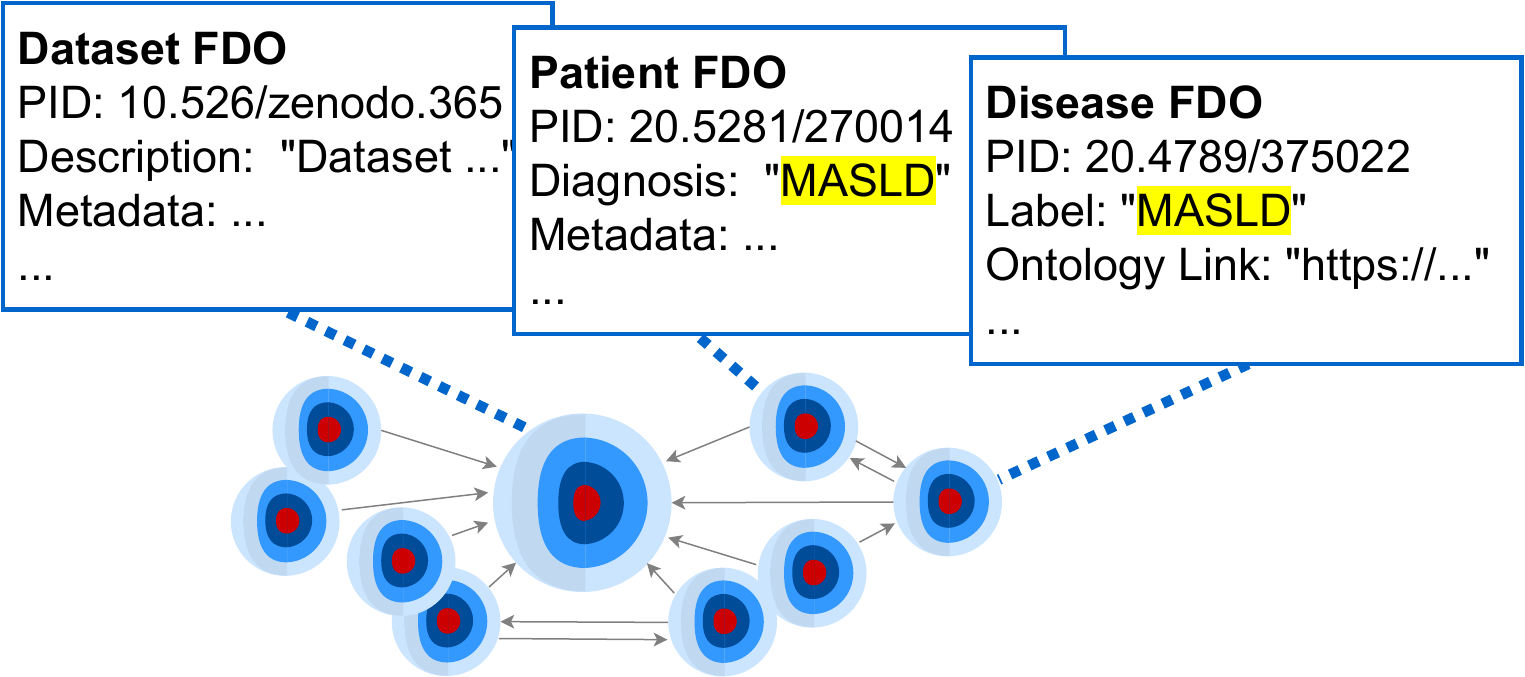}}
\caption{A FAIR knowledge graph representation with FDOs as nodes and relationships indicating similarities between nodes such as a matching disease (highlighted in yellow). Each disease and patient respectively comprises a node.}
\label{fig:fdo_net}
\end{figure}
\subsection{Formal Model}
Our KG is based on the idea of representing nodes as FDOs, as shown in Fig.~\ref{fig:fdo_net}.
Each node contains key information such as a PID and rich metadata, which are technical prerequisites to achieve FAIRness~\cite{arnold_towards_2024}. The PID allows for resource identification and retrieval, while rich metadata provides contextual information that enhances data discoverability and interoperability. Relationships between nodes indicate interrelations between the FDOs' core data or metadata. The FAIR KG always contains a central \textit{dataset} node that provides an overview and metadata describing the entire dataset. The \textit{entity} nodes, such as disease or patient nodes (see Figure~\ref{fig:fdo_net}), are user-defined. This flexible approach allows the user to guide the graph construction process and decide which entities in the dataset should be realized as nodes in the KG.

For our KG, we adopt a property graph model. Nodes and relationships can be labeled and have properties that describe them. We decide for a property graph instead of an RDF-based graph, as the property graph structure aligns well with LLM outputs such as JSON-like structures, and the concept of graph properties allows efficient metadata integration.
To better support FDOs and enable a direct mapping from the dataset and entity representations to the graph nodes, we adapt the formal definition of a property graph data model~\cite{angles_property_2018},  modifications are highlighted in blue:
\newline

\begin{figure}[b]
\centerline{\includegraphics[width=0.5\textwidth]{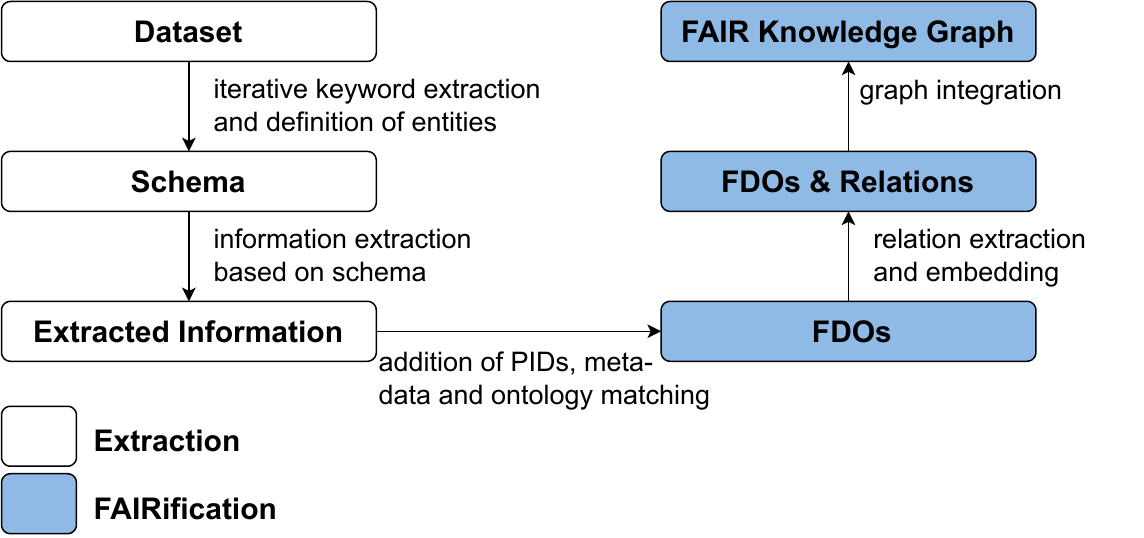}}
\caption{Graph construction stages: Based on an input dataset, a FAIR knowledge graph is constructed. The first three stages follow a schema-guided information extraction approach, while stages 3 to 6 incorporate FAIRification actions.}
\label{fig:stages}
\end{figure}

Let \boldmath\( L \) be an infinite set of labels, \( P \) an infinite set of property names, \( V \) an infinite set of atomic values, and \( T \) a finite set of datatypes. For a set \( X \), we define \( SET^{+}(X) \) as the set of all nonempty finite subsets of \( X \). Formally, our FAIR property graph is defined as a tuple:
\[
G = (\textcolor{medium_blue}{F}, E, \rho, \lambda, \sigma, \textcolor{medium_blue}{\phi})
\] where:
\begin{itemize}
    \item \( \textcolor{medium_blue}{F} \) \textcolor{medium_blue}{is a finite set of FDOs} represented as nodes;
    \item \( E \) is a finite set of edges;
    \item \( \textcolor{medium_blue}{R} \) \textcolor{medium_blue}{= \{\textit{dataset}, \textit{entity}\} is the set of node types;}
    \item \( \rho : E \rightarrow \textcolor{medium_blue}{F} \times \textcolor{medium_blue}{F} \) associates each edge with a pair of \textcolor{medium_blue}{FDOs};
    \item \( \lambda : (\textcolor{medium_blue}{F} \cup E) \rightarrow SET^{+}(L) \) associates each \textcolor{medium_blue}{FDO} with a finite set of labels;
    \item \( \sigma : (\textcolor{medium_blue}{F} \cup E) \times P \rightarrow SET^{+}(V) \) associates each \textcolor{medium_blue}{FDO} with property keys (e.g., metadata), where each property key has a finite set of values.
    \item \( \textcolor{medium_blue}{\phi : F \rightarrow R} \)  \textcolor{medium_blue}{assigns a node type to each FDO.}
\end{itemize}
In this model an FDO \( f \in F \) with \( \phi(f) = \text{\textit{dataset}} \) represents an entire dataset, an FDO \( f \in F \) with \( \phi(f) = \text{\textit{entity}} \) belongs to user-defined semantic types (e.g., \textit{patient, disease}). Edges \( (u,\, v) \in E \) can be used to represent relationships.
The formal model serves as the foundation for our FAIR KG, specifying how FDOs, their properties, and relationships are represented within a property graph framework. In the following, we outline the pipeline used to construct such a KG from a dataset. This process involves several stages, from schema definition and information extraction to FAIRification and integration into the graph database, as visualized in Fig.~\ref{fig:stages}.

\subsection{Construction of a FAIR Knowledge Graph}
Our construction pipeline is based on existing KG construction approaches~\cite{lairgi_itext2kg_2024, zhang_extract_2024, edge_local_2024}, but is adapted to the FAIR principles in stages 3 to 6.  In the following, we describe each stage, starting with the schema, which guides the information extraction process.

\begin{figure*}[t]
\centerline{\includegraphics[width=0.75\textwidth]{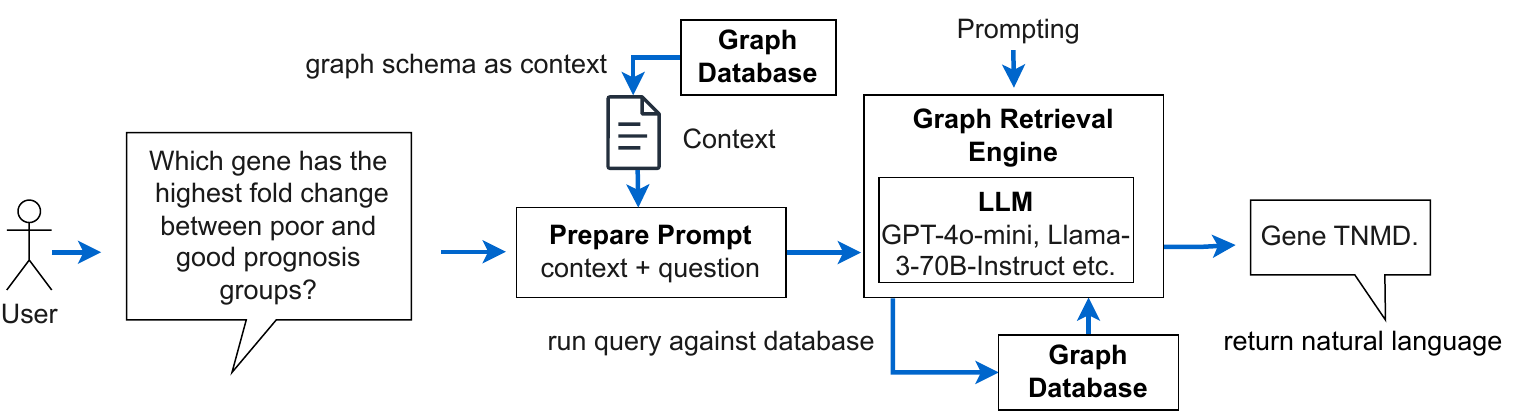}}
\caption{Graph retrieval process: The prompt is prepared based on the user question and the knowledge graph schema. The graph retrieval engine prepares the prompt, invokes the LLM to generate a database query, and executes the query against the database. Finally, the answer is returned in natural language.}
\label{fig:rag}
\end{figure*}
\subsubsection{Schema} The schema is constructed based on a fixed, basic JSON schema that is applicable to different use-cases.  Users provide an input dataset, specify the number and type of entities (e.g., \textit{patient, disease}), and add brief descriptions through a simple interface. Using few-shot prompting, an LLM then extracts relevant entity features for each user-defined entity (e.g., \textit{name, sex} for entity \textit{patient}) from the dataset to complete the initial schema. The schema aligns naturally with LLM capabilities, as it can be generated iteratively and adapted to the dataset at hand. The resulting schema acts as a blueprint, guiding the LLM to identify and extract semantically relevant information from the dataset.

\subsubsection{Extracted Information}
We follow a schema-guided knowledge extraction approach, producing structured and consistent output and allowing the user to guide the information extraction process by defining relevant entities. Depending on the nature of the source data, schema filling is
handled computationally or with LLM assistance. For structured data, we use a rule-based approach that makes use of structured data formats, maps data to schema fields and populates the schema directly. For unstructured data, the LLM is provided with the predefined schema and one document of the document collection (or dataset) at a time. The LLM is one-shot prompted to extract the relevant information from the document, fill the schema and return a filled valid JSON schema including JSON objects for each entity instance. 

\subsubsection{FAIR Digital Objects}
The extracted information is then transformed into FDOs, making use of the structured nature of the filled JSON schema. For each filled schema, a dataset FDO is created containing metadata describing the entire dataset. The metadata are identified by an LLM, which receives metadata files and is prompted to return them in a structured format, following domain-specific metadata standards. Each entity instance (e.g., the disease “pCCA”) then results in a separate entity FDO. The conversion into an FDO includes three steps:
\begin{itemize}
    \item \textbf{PID Generation:} Each entity instance and dataset is assigned a PID to ensure its findability and accessibility.
    \item \textbf{Metadata Extraction:} Using a schema-guided approach.
    \item \textbf{Ontology Term Mapping:} The schema entries are mapped to domain-specific ontology terms, enhancing interoperability.
\end{itemize}
Mapping entries to ontology terms is essential for integrating knowledge and ensuring FDO and KG interoperability. We use a flexible approach, leveraging repository APIs such as BioPortal,\footnote{\url{https://bioportal.bioontology.org}} which provides access to more than 1,200 biomedical ontologies.

\subsubsection{FAIR Digital Objects and Relations}
To enable richer knowledge representations, we extend the concept of FDOs by establishing semantic relations between them. These relations can be based on metadata similarities or core data similarities, such as the matching disease in Figure ~\ref{fig:fdo_net}.
Relations can be within a single dataset (inner-dataset) or across datasets (cross-dataset), allowing for flexible linking of related entities. Once identified, the relations are added to the schema, resulting in an enriched representation that captures both the content and the connections between FDOs.

\subsubsection{FAIR Knowledge Graph}
The final step is the integration of the FDOs and their relations into a graph representation, forming the FAIR KG. Each FDO becomes a node and each relation becomes an edge connecting related nodes. Every node in the graph
has a set of properties, including a PID, associated metadata, ontology terms, and other domain-specific attributes. 
Due to the structured nature of FDOs, KG integration is a computationally straightforward mapping process. The
resulting KG is stored in a property graph database, which enables efficient querying, exploration, and integration of knowledge. The use of a standardized graph query language (e.g., Cypher or SPARQL) ensures that the data can be accessed~\cite{brandizi_towards_2018}, supporting a wide range of analytical and semantic use cases.

\subsection{FAIR Graph Retrieval-Augmented Generation}
The graph retrieval process is visualized in Fig.~\ref{fig:rag}. To retrieve graph data, the graph guides the retrieval task, including information about nodes, edges, and their properties. A prompt is prepared incorporating the context (graph schema) and the user question. The prompt is injected to an LLM to generate a graph database query in a graph database language such as Cypher~\footnote{\url{https://neo4j.com/docs/getting-started/cypher/}} or SPARQL~\footnote{\url{https://www.w3.org/TR/sparql11-query/}}. In comparison to embedding-based similarity searches for data retrieval, query-based retrieval allows for high flexibility and can be easily adopted to various use cases. The database results are interpreted and translated into natural language, which is then returned to the user. In addition to an answer in natural language, the original graph database query and the retrieved FDOs are shown to the user, including a PID and full metadata.
\begin{table*}[t]
\caption{FAIR compliance assessment of the dataset according to FAIR indicators F1 to R1.3.}
\centering
\begin{tabular}{|l|p{10.5cm}|c|c|}
\hline
\textbf{FAIR Indicator} & \textbf{Criterion} & \textbf{FAIR-aware} & \textbf{Baseline} \\
\hline
\textbf{F1} & (Meta)data are assigned a globally unique and persistent identifier. & Yes & No\\
\textbf{F2} & Data are described with rich metadata. & Yes & No\\
\textbf{F3} & Metadata clearly include the identifier of the data they describe. & Yes & No\\
\textbf{F4} & (Meta)data are registered or indexed in a searchable resource. & Yes & Partial\\
\hline
\textbf{A1} & (Meta)data are retrievable by their identifier using a standardized communication protocol. & Yes & Yes\\
\textbf{A1.1} & The protocol is open, free, and universally implementable. & Yes & Yes\\
\textbf{A1.2} & The protocol allows authentication and authorization, where necessary. & Yes & Yes\\
\textbf{A2} & Metadata are accessible even when the data are no longer available. & Yes & No\\
\hline
\textbf{I1} & (Meta)data use a formal, accessible, shared language for knowledge representation. & Yes & Partial\\
\textbf{I2} & (Meta)data use vocabularies that follow FAIR principles. & Yes & No\\
\textbf{I3} & (Meta)data include qualified references to other (meta)data. & Yes & No\\
\hline
\textbf{R1} & (Meta)data are richly described with relevant attributes. & Yes & No\\
\textbf{R1.1} & (Meta)data are released with a clear and accessible data usage license. & No & No\\
\textbf{R1.2} & (Meta)data are associated with detailed provenance. & Yes & No\\
\textbf{R1.3} & (Meta)data meet domain-relevant community standards. & Yes & Partial\\
\hline
\end{tabular}
\label{tab:fair-f1-r13}
\end{table*}
\section{Experiments and Results}
\label{sec:experiment}
We evaluate FAIR GraphRAG on a biomedical dataset to show the advantages of our approach compared to a Non-FAIR GraphRAG approach.

\subsection{Experimental Settings}

\subsubsection{Dataset}
The RNA sequencing data reported in this study have been deposited in the Gene Expression Omnibus (GEO) under accession GEO Series GSE280797~\footnote{\url{https://www.ncbi.nlm.nih.gov/geo/query/acc.cgi}} (Peking University Cancer Hospital and Institute, submitted October 31, 2024; last updated January 30, 2025). These data correspond to the study “Screening and molecular mechanism research on bile microRNAs associated with chemotherapy efficacy in perihilar cholangiocarcinoma”~\cite{fu_screening_2024}.
The data is derived from bile samples of patients with the rare liver cancer perihilar cholangiocarcinoma (pCCA). Samples were collected from 8 patients prior to chemotherapy treatment, as part of a retrospective clinical study. 
We used processed data in a tabular format, including gene annotations, expression values (FPKM), fold changes, and statistical significance. GEO enforces MIAME~\footnote{\url{https://www.ncbi.nlm.nih.gov/geo/info/MIAME.html}} compliant metadata to ensure consistency across platform, sample, and experiment annotations. The dataset contains MIAME compliant metadata on the dataset and its samples.

\subsubsection{Evaluation Metrics}
We evaluate the adherence to the FAIR principles by assessing each of the four dimensions \textit{findability}, \textit{accessibility}, \textit{interoperability}, and \textit{reusability} answering questions F1 to R1.3 as taken from~\cite{wilkinson_fair_2016} on each dimension, listed in Table~\ref{tab:fair-f1-r13}. Furthermore, we assess \textit{accuracy}, \textit{coverage}, and \textit{explainability}. Accuracy and coverage metrics ensure that the system reliably returns correct answers across a set of representative queries. The query set includes 22 general questions (e.g., “Which gene shows the largest difference in expression (fold change) between Group A and Group B?”), 10 metadata-specific questions (e.g., “What is the sequencing instrument used for this study?”), and 10 ontology-related questions (e.g., “Provide the GO-BP ontology term linked to the gene CFLAR.”).  While LLMs are used for graph construction and graph database query generation, limiting the system's explainability, we assess whether the system supports tracing back data sources, identifiers, and provenance details. In our use case, a fully explainable answer contains: (1) the Cypher query used, (2) the graph nodes returned, and (3) FAIR features (PID, metadata, ontology links). Criteria are weighted equally in the evaluation. While our evaluation focuses on technical performance and FAIR compliance, we do not include clinical evaluation criteria such as validation by medical experts, as this is outside the scope of this technical proof of concept.

\subsubsection{Baseline}
We introduce a Non-FAIR GraphRAG approach, with an underlying Non-FAIR KG that was constructed following our KG construction pipeline. FAIRification steps, such as assigning PIDs, metadata, and ontology terms, were skipped.

\subsubsection{Implementation Details}
We construct the KG using both OpenAI’s gpt-4o-mini and Meta’s open-source model Llama-3.3-70B-Instruct, and observe that both produce the same KG. Model inference is performed via Azure Services, protecting the data from proprietary use. To fit the LLM context window and minimize cost, we use the first 80 rows of the RNA sequencing dataset. We define \textit{pathway} and \textit{GO-BP} as separate user-defined entities, modeling them as nodes to enable semantic linking across genes. To make PIDs resolvable, we simulate dataset registration in the Zenodo Sandbox\footnote{\url{https://sandbox.zenodo.org/}}, as we are not the dataset owners and cannot register data at the main Zenodo repository\footnote{\url{https://zenodo.org/}}. The resulting KG (1038 nodes: 1 dataset, 80 gene, 854 GO-BP, 103 pathway nodes) is stored in Neo4j\footnote{\url{https://neo4j.com/}}. For RAG, we use LangChain\footnote{\url{https://www.langchain.com/}} to generate Cypher queries from the KG schema, as shown in Fig.~\ref{fig:rag}.
The source code is available as open-source on GitHub \cite{Fluh_FAIR_Knowledge_Graph_2025, Fluh_FAIR_RAG_Interface_2025}.

\begin{table*}[t]
\centering
\caption{Comparison of accuracy on different question categories}
\begin{tabular}{|l|l|c|c|c|c|}
\hline
\textbf{Model}       & \textbf{System} & \textbf{General Acc} & \textbf{Metadata Acc} & \textbf{Ontology Acc} & \textbf{Overall Acc} \\
\hline
gpt-4o-mini          & FAIR-aware            & 95.45\%                   & 90\%                    & 90\%                    & 92.86\%                   \\
Llama-3.3-70B        & FAIR-aware            & 45.45\%                   & 20\%                    & 20\%                    & 33.33\%                   \\
gpt-4o-mini          & Baseline        & 81.82\%                   & 0\%                    & 0\%                    & 42.86\%                   \\
Llama-3.3-70B        & Baseline        & 36.36\%                   & 0\%                    & 0\%                    & 19.05\%                   \\
\hline
\end{tabular}
\label{tab:acc_comparison}
\end{table*}

\begin{table}[t]
\centering
\caption{Comparison of answer accuracy, coverage, and explainability}
\begin{tabular}{|l|l|c|c|c|}
\hline
\textbf{Model}       & \textbf{System} & \textbf{ Acc. (\%)} & \textbf{Cov. (\%)} & \textbf{Expl. (\%)}$^{\mathrm{a}}$ \\
\hline
gpt-4o-mini          & FAIR-aware            & 92.86\%                   & 92.86\%                    & 100\%                    \\
Llama-3.3-70B        & FAIR-aware            & 33.33\%                   & 35.71\%                    & 100\%                    \\
gpt-4o-mini          & Baseline        & 42.86\%                   & 42.86\%                    & 66.66\%                    \\
Llama-3.3-70B        & Baseline        & 19.05\%                   & 21.43\%                    & 66.66\%                    \\
\hline
\multicolumn{5}{l}{$^{\mathrm{a}}$Cypher queries are generated by an LLM, preventing full explainability.}
\end{tabular}
\label{tab:rag_comparison}
\end{table}

\subsection{Main Results}
\subsubsection{FAIR Results}
We assess FAIR compliance by answering questions F1 to R1.3, the results are summarized in Table~\ref{tab:fair-f1-r13}. Our GraphRAG approach meets all FAIR criteria except criterion R1.1, as the dataset lacks licensing information. The baseline approach shows poor results for \textit{findability, interoperability,} and \textit{reusability} criteria due to missing PIDs, metadata, and ontology term mapping. Both approaches are accessible through HTTP(S) and Bolt protocols, and support authentication and role-based access, as the underlying KGs are stored in Neo4j. These results demonstrate that our GraphRAG approach achieves significantly better FAIR compliance than the baseline, as reflected in Table~\ref{tab:fair-f1-r13}.
\subsubsection{Question Answering Performance}
The performance results are summarized in Table~\ref{tab:acc_comparison} and Table~\ref{tab:rag_comparison}. Our FAIR GraphRAG approach, when used with gpt-4o-mini, achieves the highest overall accuracy (92.86\%), coverage (92.86\%) and explainability (100\%) in all types of questions. The same model applied to the Non-FAIR KG achieves an overall accuracy and coverage of 42.86\%, with reduced explainability.
Similarly, Llama-3.3-70B demonstrates improved performance in the FAIR setting compared to the Non-FAIR baseline, but still lags behind gpt-4o-mini. The Non-FAIR system fails to answer metadata and ontology-specific questions, resulting in 0\% accuracy in these categories (Table~\ref{tab:acc_comparison}).
For both models, the FAIR KG improves the system's ability to answer complex queries, especially when involving metadata and ontology links.
The results demonstrate that a FAIR-aligned KG enhances the quality of answers in this biomedical question answering use case.

\subsection{Experiment Analysis}
\begin{figure}[b]
\centerline{\includegraphics[width=0.5\textwidth]{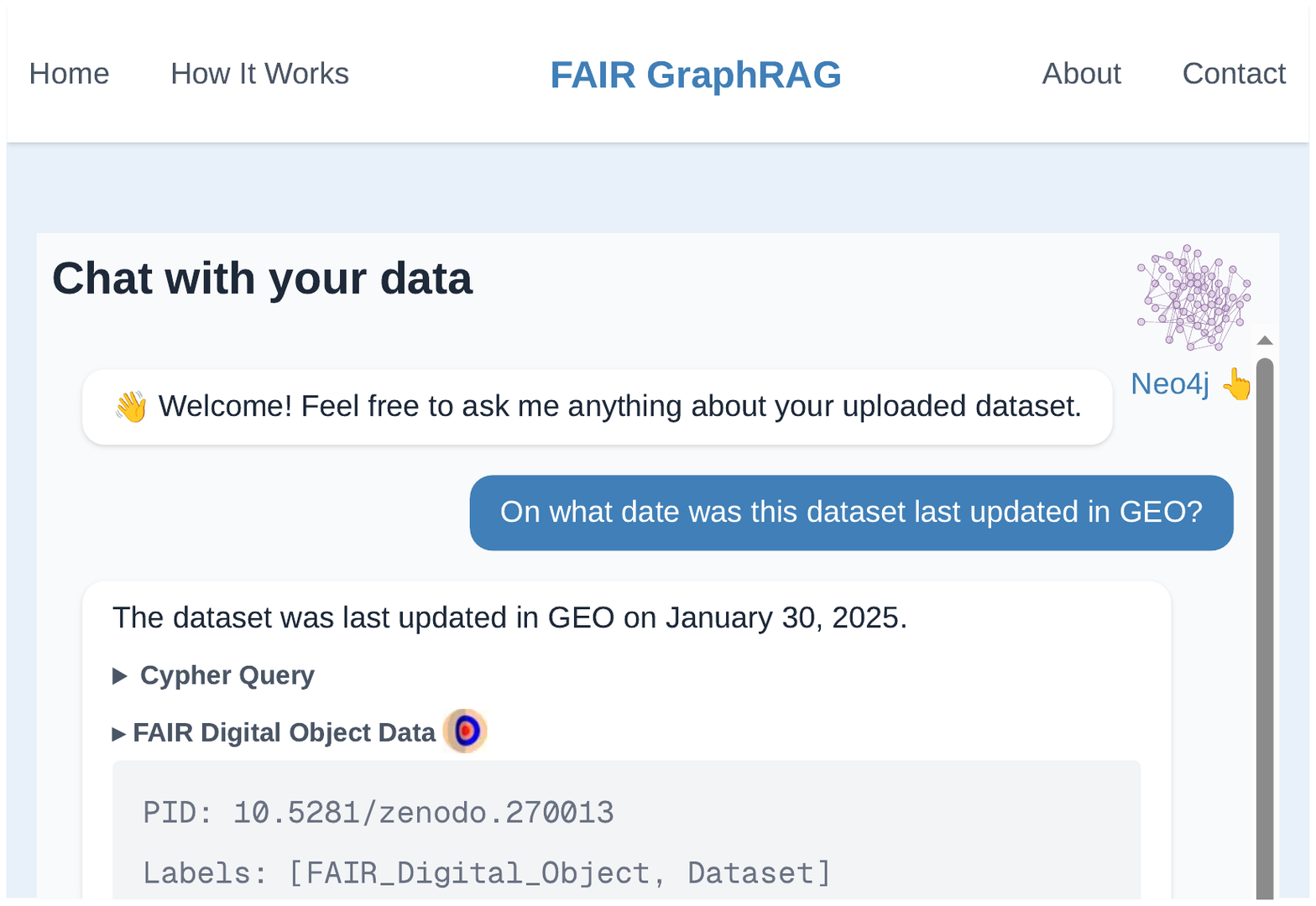}}
\caption{Chat Interface: The chat interface enables knowledge graph querying in natural language and provides details on the relevant Fair Digital Object.}
\label{fig:chat_ui}
\end{figure}

As shown in Table~\ref{tab:fair-f1-r13}, our GraphRAG approach meets all FAIR criteria except licensing by including Zenodo PIDs, a shared vocabulary and links to biomedical ontologies. The metadata is MIAME-compliant and exceeds the detail offered by general metadata vocabulary such as Dublin Core Metadata Terms (DCMT)~\footnote{\url{https://www.dublincore.org/specifications/dublin-core/dces/}}.
Table~\ref{tab:acc_comparison} reports accuracy by question type. The FAIR system using gpt-4o-mini demonstrates robust performance on all question types, with rare failures on complex questions (e.g., exceeding the token limit). The baseline system achieves lower accuracy, especially for metadata and ontology questions, due to missing KG properties and links.
Llama-3.3-70B underperforms on both systems, often generating invalid Cypher queries, which return no results.
For FAIR and Non-FAIR settings, gpt-4o-mini's accuracy matches its coverage, indicating that it only answers when confident, with no incorrect results. 
Explainability is highest in the FAIR system, as in addition to Cypher queries and node data, it provides ontology links, shared vocabulary, and PIDs, facilitating full provenance and traceability. In contrast, the Non-FAIR system lacks this semantic richness.  Fig.~\ref{fig:chat_ui} shows the chat interface with an example question, a Cypher query and dataset FDO. The resolvable PID is demonstrated using Zenodo Sandbox.

\section{Conclusion}
\label{sec:conclusion}
In this work, we introduce FAIR GraphRAG, a novel approach that integrates FAIR Digital Objects into graph-based RAG, advancing both the FAIRification and semantic interoperability of domain-specific knowledge graphs. Leveraging LLMs, FAIR GraphRAG supports schema construction and automated extraction of content and metadata. Our experiments on biomedical data demonstrate that FAIR GraphRAG significantly improves findability, interoperability, reusability, and explainability over baseline approaches, as seen in both FAIR compliance and question-answering metrics. By employing standardized metadata, PIDs, and ontology term mapping, FAIR GraphRAG achieves transparent and accurate results. The approach is adaptable to other fields, supporting the broader adoption of FAIR data principles in retrieval-augmented generation systems. Our framework supports both open-source and proprietary LLMs, enabling flexible and privacy-conscious deployment that complies with ethical and regulatory standards. In future work, we will further explore automated FAIRification and enhancing semantic interoperability with both internal and external resources. This will involve advancing methods for improved metadata linking and ontology term mapping.

\section*{Acknowledgment}
This work was funded under the Excellence Strategy of the Federal Government and the Länder (PFExC005 -- Computational ecosystem for clinical applications of organ crosstalk), and the Deutsche Forschungsgemeinschaft (DFG, German Research Foundation) under Germany’s Excellence Strategy -- EXC2023 Internet of Production -- 390621612.

\bibliographystyle{IEEEtran}
\bibliography{bibliography}

\begin{thebibliography}{10}
\providecommand{\url}[1]{#1}
\csname url@samestyle\endcsname
\providecommand{\newblock}{\relax}
\providecommand{\bibinfo}[2]{#2}
\providecommand{\BIBentrySTDinterwordspacing}{\spaceskip=0pt\relax}
\providecommand{\BIBentryALTinterwordstretchfactor}{4}
\providecommand{\BIBentryALTinterwordspacing}{\spaceskip=\fontdimen2\font plus
\BIBentryALTinterwordstretchfactor\fontdimen3\font minus \fontdimen4\font\relax}
\providecommand{\BIBforeignlanguage}[2]{{%
\expandafter\ifx\csname l@#1\endcsname\relax
\typeout{** WARNING: IEEEtran.bst: No hyphenation pattern has been}%
\typeout{** loaded for the language `#1'. Using the pattern for}%
\typeout{** the default language instead.}%
\else
\language=\csname l@#1\endcsname
\fi
#2}}
\providecommand{\BIBdecl}{\relax}
\BIBdecl

\bibitem{pan_large_2023}
\BIBentryALTinterwordspacing
J.~Z. Pan, S.~Razniewski, J.-C. Kalo, S.~Singhania, J.~Chen, S.~Dietze, H.~Jabeen, J.~Omeliyanenko, W.~Zhang, M.~Lissandrini, R.~Biswas, G.~de~Melo, A.~Bonifati, E.~Vakaj, M.~Dragoni, and D.~Graux, ``Large {Language} {Models} and {Knowledge} {Graphs}: {Opportunities} and {Challenges},'' Aug. 2023, arXiv:2308.06374 [cs]. [Online]. Available: \url{http://arxiv.org/abs/2308.06374}
\BIBentrySTDinterwordspacing

\bibitem{peng_graph_2024}
B.~Peng, Y.~Zhu, Y.~Liu, X.~Bo, H.~Shi, C.~Hong, Y.~Zhang, and S.~Tang, ``Graph retrieval-augmented generation: A survey,'' Sep. 2024, arXiv preprint arXiv:2408.08921. Available: \href{https://arxiv.org/abs/2408.08921}{arxiv.org/abs/2408.08921}.

\bibitem{collarana_graphrag_2025}
\BIBentryALTinterwordspacing
D.~Collarana, C.~I. Pack, Y.-Y. Liao, M.~Flüh, J.~Lehmkuhl, A.~Nageri, A.~Graß, M.~Busch, P.~Das, L.~Dingels, S.~Decker, and C.~Beecks, ``Graph rag in the wild: Insights and best practices from real-world applications,'' \emph{Semantic Web Journal}, 2025, under review. [Online]. Available: \url{https://www.semantic-web-journal.net/content/graph-rag-wild-insights-and-best-practices-real-world-applications}
\BIBentrySTDinterwordspacing

\bibitem{yu_evaluation_2025}
H.~Yu, A.~Gan, K.~Zhang, S.~Tong, Q.~Liu, and Z.~Liu, ``\BIBforeignlanguage{en}{Evaluation of {Retrieval}-{Augmented} {Generation}: {A} {Survey}},'' in \emph{\BIBforeignlanguage{en}{Big {Data}}}, W.~Zhu, H.~Xiong, X.~Cheng, L.~Cui, Z.~Dou, J.~Dong, S.~Pang, L.~Wang, L.~Kong, and Z.~Chen, Eds.\hskip 1em plus 0.5em minus 0.4em\relax Singapore: Springer Nature, 2025, pp. 102--120.

\bibitem{zhao_retrieval-augmented_2024}
\BIBentryALTinterwordspacing
P.~Zhao, H.~Zhang, Q.~Yu, Z.~Wang, Y.~Geng, F.~Fu, L.~Yang, W.~Zhang, J.~Jiang, and B.~Cui, ``Retrieval-{Augmented} {Generation} for {AI}-{Generated} {Content}: {A} {Survey},'' Jun. 2024, arXiv:2402.19473 [cs]. [Online]. Available: \url{http://arxiv.org/abs/2402.19473}
\BIBentrySTDinterwordspacing

\bibitem{ji_survey_2022}
\BIBentryALTinterwordspacing
S.~Ji, S.~Pan, E.~Cambria, P.~Marttinen, and P.~S. Yu, ``A {Survey} on {Knowledge} {Graphs}: {Representation}, {Acquisition} and {Applications},'' \emph{IEEE Transactions on Neural Networks and Learning Systems}, vol.~33, no.~2, pp. 494--514, Feb. 2022, arXiv:2002.00388 [cs]. [Online]. Available: \url{http://arxiv.org/abs/2002.00388}
\BIBentrySTDinterwordspacing

\bibitem{de_smedt_fair_2020}
\BIBentryALTinterwordspacing
K.~De~Smedt, D.~Koureas, and P.~Wittenburg, ``\BIBforeignlanguage{en}{{FAIR} {Digital} {Objects} for {Science}: {From} {Data} {Pieces} to {Actionable} {Knowledge} {Units}},'' \emph{\BIBforeignlanguage{en}{Publications}}, vol.~8, no.~2, p.~21, Jun. 2020, number: 2 Publisher: Multidisciplinary Digital Publishing Institute. [Online]. Available: \url{https://www.mdpi.com/2304-6775/8/2/21}
\BIBentrySTDinterwordspacing

\bibitem{brandizi_towards_2018}
M.~Brandizi, A.~Singh, C.~Rawlings, and K.~Hassani-Pak, ``\BIBforeignlanguage{eng}{Towards {FAIRer} {Biological} {Knowledge} {Networks} {Using} a {Hybrid} {Linked} {Data} and {Graph} {Database} {Approach}},'' \emph{\BIBforeignlanguage{eng}{Journal of Integrative Bioinformatics}}, vol.~15, no.~3, p. 20180023, Aug. 2018.

\bibitem{pestryakova_covidpubgraph_2022}
\BIBentryALTinterwordspacing
S.~Pestryakova, D.~Vollmers, M.~A. Sherif, S.~Heindorf, M.~Saleem, D.~Moussallem, and A.-C.~N. Ngomo, ``\BIBforeignlanguage{en}{{CovidPubGraph}: {A} {FAIR} {Knowledge} {Graph} of {COVID}-19 {Publications}},'' \emph{\BIBforeignlanguage{en}{Scientific Data}}, vol.~9, no.~1, p. 389, Jul. 2022, publisher: Nature Publishing Group. [Online]. Available: \url{https://www.nature.com/articles/s41597-022-01298-2}
\BIBentrySTDinterwordspacing

\bibitem{sherif_ingridkg_2023}
M.~A. Sherif, A.~A.~M. da~Silva, S.~Pestryakova, A.~F. Ahmed, S.~Niemann, and A.-C.~N. Ngomo, ``\BIBforeignlanguage{eng}{{INGRIDKG}: {A} {FAIR} {Knowledge} {Graph} of {Graffiti}},'' \emph{\BIBforeignlanguage{eng}{Scientific Data}}, vol.~10, no.~1, p. 318, May 2023.

\bibitem{vogt_fair_2023}
L.~Vogt, ``{FAIR} knowledge graphs with semantic units: A prototype,'' Nov. 2023, arXiv preprint arXiv:2311.04761. Available: \href{https://arxiv.org/abs/2311.04761}{arxiv.org/abs/2311.04761}.

\bibitem{wittenburg_digital_2019}
P.~Wittenburg, ``\BIBforeignlanguage{en}{Digital objects as drivers towards convergence in data infrastructures},'' 2019, available: \href{https://b2share.eudat.eu/records/b605d85809ca45679b110719b6c6cb11}{b2share.eudat.eu/records/...}

\bibitem{wilkinson_fair_2016}
\BIBentryALTinterwordspacing
M.~D. Wilkinson, M.~Dumontier, I.~J. Aalbersberg, G.~Appleton, M.~Axton, A.~Baak, N.~Blomberg, J.-W. Boiten, L.~B. da~Silva~Santos, P.~E. Bourne, J.~Bouwman, A.~J. Brookes, T.~Clark, M.~Crosas, I.~Dillo, O.~Dumon, S.~Edmunds, C.~T. Evelo, R.~Finkers, A.~Gonzalez-Beltran, A.~J.~G. Gray, P.~Groth, C.~Goble, J.~S. Grethe, J.~Heringa, P.~A.~C. ’t Hoen, R.~Hooft, T.~Kuhn, R.~Kok, J.~Kok, S.~J. Lusher, M.~E. Martone, A.~Mons, A.~L. Packer, B.~Persson, P.~Rocca-Serra, M.~Roos, R.~van Schaik, S.-A. Sansone, E.~Schultes, T.~Sengstag, T.~Slater, G.~Strawn, M.~A. Swertz, M.~Thompson, J.~van~der Lei, E.~van Mulligen, J.~Velterop, A.~Waagmeester, P.~Wittenburg, K.~Wolstencroft, J.~Zhao, and B.~Mons, ``\BIBforeignlanguage{en}{The {FAIR} {Guiding} {Principles} for scientific data management and stewardship},'' \emph{\BIBforeignlanguage{en}{Scientific Data}}, vol.~3, no.~1, p. 160018, Mar. 2016, number: 1 Publisher: Nature Publishing Group. [Online]. Available: \url{https://www.nature.com/articles/sdata201618}
\BIBentrySTDinterwordspacing

\bibitem{schwardmann_digital_2020}
U.~Schwardmann, ``\BIBforeignlanguage{en-US}{Digital objects – fair digital objects: Which services are required?}'' \emph{\BIBforeignlanguage{en-US}{Data Science Journal}}, vol.~19, no.~1, Apr. 2020, available: \href{https://datascience.codata.org/articles/10.5334/dsj-2020-015}{datascience.codata.org/articles/10.5334/dsj-2020-015}.

\bibitem{queralt-rosinach_applying_2022}
N.~Queralt-Rosinach, R.~Kaliyaperumal, C.~H. Bernabé, Q.~Long, S.~A. Joosten, H.~J. van~der Wijk, E.~L.~A. Flikkenschild, K.~Burger, A.~Jacobsen, B.~Mons, M.~Roos, {BEAT-COVID Group}, and {COVID-19 LUMC Group}, ``\BIBforeignlanguage{eng}{Applying the {FAIR} principles to data in a hospital: challenges and opportunities in a pandemic},'' \emph{\BIBforeignlanguage{eng}{Journal of Biomedical Semantics}}, vol.~13, no.~1, p.~12, Apr. 2022.

\bibitem{vogt_towards_2024}
L.~Vogt, C.~Biniossek, D.~Betz, and M.~Stocker, ``Towards machine-actionable scientific knowledge as {FAIR} {Digital} {Objects},'' Mar. 2024.

\bibitem{flynn_linked_2022}
\BIBentryALTinterwordspacing
A.~Flynn, M.~Conte, P.~Boisvert, R.~Richesson, Z.~Landis-Lewis, and C.~Friedman, ``\BIBforeignlanguage{en}{Linked {Metadata} for {FAIR} {Digital} {Objects} {Carrying} {Computable} {Knowledge}},'' \emph{\BIBforeignlanguage{en}{Research Ideas and Outcomes}}, vol.~8, p. e94438, Oct. 2022, publisher: Pensoft Publishers. [Online]. Available: \url{https://riojournal.com/article/94438/}
\BIBentrySTDinterwordspacing

\bibitem{cappiello_52_2024}
\BIBentryALTinterwordspacing
C.~Cappiello, M.~E. Vidal, S.~Geisler, A.~Iglesias-Molina, D.~Van~Assche, D.~Chaves-Fraga, A.~Dimou, I.~Celino, A.~Rula, and M.~Lenzerini, ``5.2 {Quality}-aware {Knowledge} {Graph} {Construction},'' \emph{Are Knowledge Graphs Ready for the Real World? Challenges and Perspective}, vol.~20, p.~47, 2024. [Online]. Available: \url{https://lirias.kuleuven.be/retrieve/777295#page=49}
\BIBentrySTDinterwordspacing

\bibitem{vogt_semantic_2024}
\BIBentryALTinterwordspacing
L.~Vogt, T.~Kuhn, and R.~Hoehndorf, ``Semantic units: organizing knowledge graphs into semantically meaningful units of representation,'' \emph{Journal of Biomedical Semantics}, vol.~15, no.~1, p.~7, May 2024. [Online]. Available: \url{https://doi.org/10.1186/s13326-024-00310-5}
\BIBentrySTDinterwordspacing

\bibitem{sun_docs2kg_2024}
Q.~Sun, Y.~Luo, W.~Zhang, S.~Li, J.~Li, K.~Niu, X.~Kong, and W.~Liu, ``{Docs2KG}: Unified knowledge graph construction from heterogeneous documents assisted by large language models,'' Jun. 2024, arXiv preprint arXiv:2406.02962. Available: \href{https://arxiv.org/abs/2406.02962}{arxiv.org/abs/2406.02962}. Project: \href{https://docs2kg.ai4wa.com}{docs2kg.ai4wa.com}.

\bibitem{lairgi_itext2kg_2024}
\BIBentryALTinterwordspacing
Y.~Lairgi, L.~Moncla, R.~Cazabet, K.~Benabdeslem, and P.~Cléau, ``{iText2KG}: {Incremental} {Knowledge} {Graphs} {Construction} {Using} {Large} {Language} {Models},'' Sep. 2024, arXiv:2409.03284 [cs]. [Online]. Available: \url{http://arxiv.org/abs/2409.03284}
\BIBentrySTDinterwordspacing

\bibitem{zhang_extract_2024}
\BIBentryALTinterwordspacing
B.~Zhang and H.~Soh, ``Extract, {Define}, {Canonicalize}: {An} {LLM}-based {Framework} for {Knowledge} {Graph} {Construction},'' Oct. 2024, arXiv:2404.03868 [cs]. [Online]. Available: \url{http://arxiv.org/abs/2404.03868}
\BIBentrySTDinterwordspacing

\bibitem{edge_local_2024}
\BIBentryALTinterwordspacing
D.~Edge, H.~Trinh, N.~Cheng, J.~Bradley, A.~Chao, A.~Mody, S.~Truitt, and J.~Larson, ``From {Local} to {Global}: {A} {Graph} {RAG} {Approach} to {Query}-{Focused} {Summarization},'' Apr. 2024, arXiv:2404.16130. [Online]. Available: \url{http://arxiv.org/abs/2404.16130}
\BIBentrySTDinterwordspacing

\bibitem{arnold_towards_2024}
B.~T. Arnold, J.~Theissen-Lipp, D.~Collarana, C.~Lange-Bever, S.~Geisler, E.~Curry, and S.~J. Decker, \emph{\BIBforeignlanguage{English}{Towards {Enabling} {FAIR} {Dataspaces} {Using} {Large} {Language} {Models}}}.\hskip 1em plus 0.5em minus 0.4em\relax CEUR-WS, 2024, medium: online Meeting Name: 21. Extended Semantic Web Conference.

\bibitem{angles_property_2018}
\BIBentryALTinterwordspacing
R.~Angles, ``The {Property} {Graph} {Database} {Model},'' in \emph{Proceedings of the 12th {Alberto} {Mendelzon} {International} {Workshop} on {Foundations} of {Data} {Management}, {Cali}, {Colombia}, {May} 21-25, 2018}, ser. {CEUR} {Workshop} {Proceedings}, D.~Olteanu and B.~Poblete, Eds., vol. 2100.\hskip 1em plus 0.5em minus 0.4em\relax CEUR-WS.org, 2018. [Online]. Available: \url{https://ceur-ws.org/Vol-2100/paper26.pdf}
\BIBentrySTDinterwordspacing

\bibitem{fu_screening_2024}
\BIBentryALTinterwordspacing
S.~Fu, H.~Du, Y.~Dai, K.~Zheng, G.~Cao, L.~Xu, Y.~Zhong, C.~Niu, Y.~Kong, and X.~Wang, ``Screening and molecular mechanism research on bile {microRNAs} associated with chemotherapy efficacy in perihilar cholangiocarcinoma,'' \emph{iScience}, vol.~27, no.~12, p. 111437, Dec. 2024. [Online]. Available: \url{https://www.sciencedirect.com/science/article/pii/S2589004224026622}
\BIBentrySTDinterwordspacing

\bibitem{Fluh_FAIR_Knowledge_Graph_2025}
M.~Flüh, ``{FAIR Knowledge Graph Construction},'' \url{https://github.com/FAIR-GraphRAG/CSV2FAIR_KG}, 2025, accessed: Sep. 19, 2025.

\bibitem{Fluh_FAIR_RAG_Interface_2025}
------, ``{FAIR Retrieval-Augmented Generation Interface},'' \url{https://github.com/FAIR-GraphRAG/RAG-interface}, Sep. 2025, accessed: Sep. 19, 2025.

\end{thebibliography}
\vspace{12pt}
\end{document}